\documentclass[pra, twocolumn,  superscriptaddress]{revtex4}
\usepackage{amssymb}
\usepackage{amsmath}
\usepackage{graphicx}
\usepackage{dcolumn}
\usepackage{color}
\usepackage{bm}
\usepackage{subfigure}
\usepackage{amsfonts}

\begin{document}
\preprint{APS/123-QED}
\title{ Compiling  universal  quantum circuits }
\author{Rahul Pratap Singh }
\affiliation{Department of Physical Sciences, Indian Institute of Science Education and Research
Kolkata, Mohanpur 741246, West Bengal, India}
\affiliation{Department of Physics, School of Science and Humanities, Nazarbayev University, 53 Kabanbay Batyr Avenue, Nur-Sultan 010000, Kazakhstan}
\author{ Aikaterini Mandilara}
\affiliation{Department of Physics, School of Science and Humanities, Nazarbayev University, 53 Kabanbay Batyr Avenue, Nur-Sultan 010000, Kazakhstan}
\begin{abstract}

We propose a method of compiling  that permits to identify quantum circuits  able to simulate arbitrary  $n$-qubit unitary operations via the adjustment of angles in single-qubit gates therein. The method of compiling itself extends older quantum control techniques  and  stays computationally tractable  for several qubits.
As an application we identify compiling universal circuits for $3$, $4$ and $5$ qubits consisting of $16$, $64$ and $ 256$ CNOTs  respectively.

\end{abstract}
\maketitle

\section{Introduction}

The aim of quantum compiling \cite{Kitaev, QEP, Shen} is to approximate   a given unitary operation
via  quantum circuits composed exclusively by gates drawn from   computationally \cite{Deutsch2} or efficiently universal sets \cite{Harrow}.
The demand to confine the design within specific sets of gates  is due to constrains imposed by fault-tolerance and error-correction techniques.
Throughout the years, a variety of effective methods    have been proposed for  compiling  single-qubit gates \cite{Nielsen, Fowler, Svore, Booth, SvoreII, Horsman, Mosca, mand}, nevertheless  very little progress has been done towards  higher dimensions because of the exponentially increasing difficulty of the problem.

A closely related problem \cite{ QEP, Deutsch} arises  if   single-qubit operations with free parameters are added  to the set of available  gates.
This problem   provides with exact results in compiling, is more tractable  in higher-dimensions \cite{QEP, Salomaa1, Salomaa3, Vidal, Markov} and naturally results to less lengthy sequences of gates than the original problem does. A solution to this problem  does not provide fault-tolerant circuits but  is of   potential  interest  for  the simulation of  many-body quantum dynamics \cite{Sanders} on quantum circuits. 
The   task  is mostly known in the literature as the  efficient gate decomposition problem since  historically this has been formulated \cite{Deutsch} and solved
by combining two-level unitary matrices decompositions \cite{Reck} with  Gray codes \cite{Deutsch, QEP, Salomaa1, Salomaa3}. An alternative way of approaching the problem using  quantum multiplexors has been suggested in \cite{Markov},  providing the lowest counts on two-qubit gates thus far, i.e. $\frac{23}{48} 4^n - \frac{3}{ 2} 2^n + \frac{4} {3}$ CNOTs, where $n$ the number of qubits. Finally, 
more recently there are new suggestions including hybrid quantum-classical methods \cite{Kharti, Ar}, 
and approaches based on pure quantum control techniques \cite{San}.
 In  the aforementioned works though the  structure of the resulting circuit is either dictated by the outcome of the matrix decomposition \cite{Salomaa1, Salomaa3} or
by the numerical optimization \cite{Kharti}
or  by the type of the generating gate to be simulated \cite{San}. 
In this work we proceed differently, fixing  the  architecture
of the circuit and then applying a compiling method that tests its ability to compile arbitrary unitary operations and eventually tunes the local parameters of  the circuit so that this  matches a targeted operation.

More specifically, we consider circuits composed of $2^n$ \textsl{circuit units} with identical placement of two-qubit and  single-qubit gates. The architecture and the  type of  two-qubit gates in the circuit unit, can be initially determined and fixed upon convenience but the total structure should be proven efficient for compiling.
The problem of  compiling, via the adjustment of the angles in the single qubit gates, is  treated by employing and extending  quantum control techniques developed in \cite{FE}  in the context of  simulating the Floquet evolution of quantum systems. In a nutshell,  
 first   only the square root of the total parameters are adjusted so that a non-trivial unity is reached      and then departing from there, the target unitary is reached with  successive  application of  gradient descent  along a discrete path in the geometric space of unitary operators.
Regarding the scalability of our proposal. We have tested the efficiency of the compiling method   up to $5$ qubits. 
With an increase of the computational capacity and/or the sophistication of the numerical methods,   this number can be moderately increased up to $7$ qubits.


The manuscript is structured as follows. In Section~\ref{c} we present the structure of circuits under consideration and then
in Section~\ref{CM} we deploy the compiling algorithm. In Section~\ref{D} we present examples of  compiling universal circuits identified by our methods, and which are composed by CNOT two-qubit gates.

\section{The class of circuits under study \label{c}}

\begin{figure}[h] \includegraphics
[width=0.45\textwidth]
{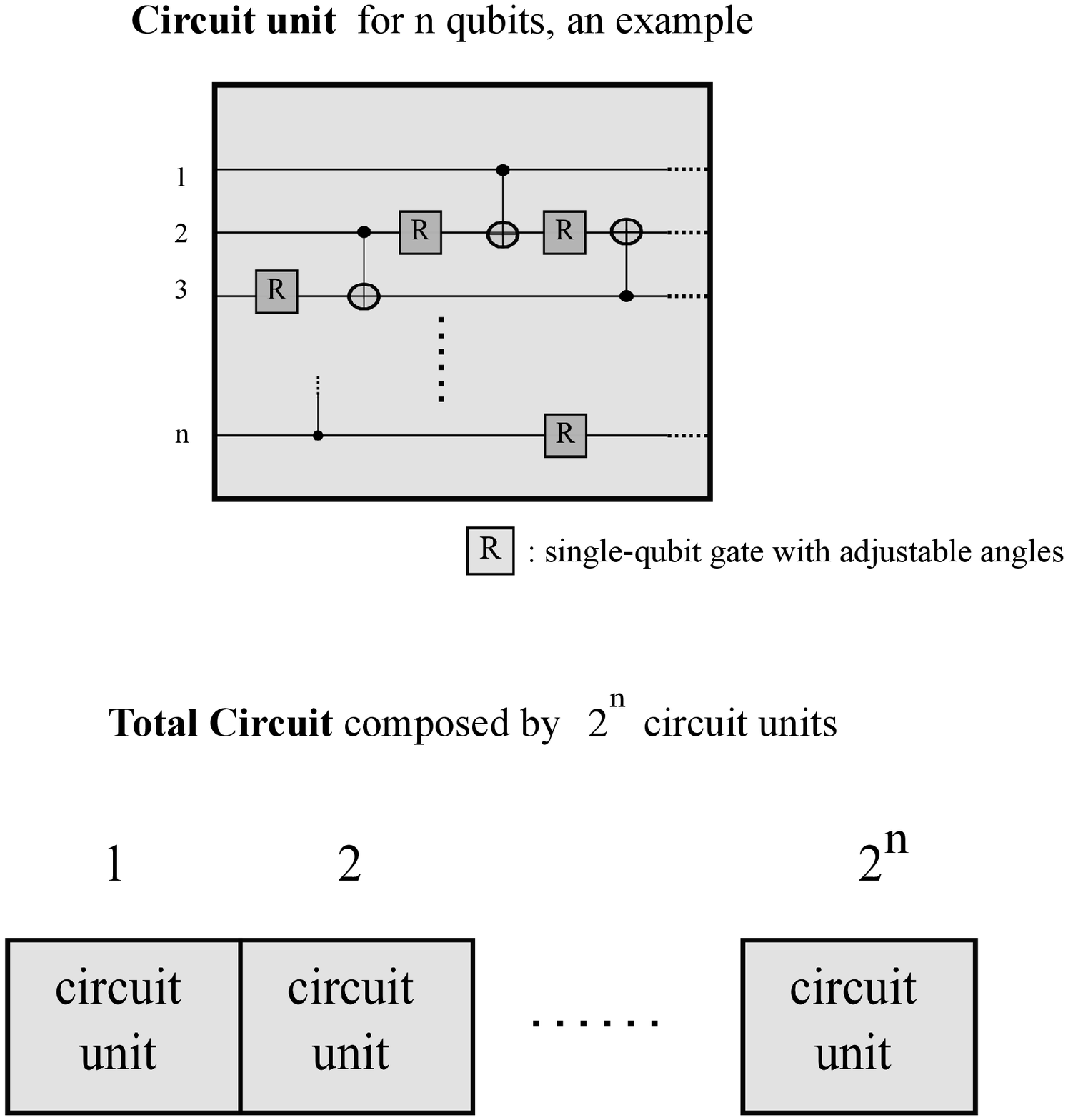}
\vspace{1cm}
\caption{ The total circuit  composed by $N=2^n$ circuit units of identical structure.
The type and placement of two-qubit gates can be arbitrary in principle. } \label{fig1}\end{figure}

The potential  is to construct circuits of $n$-qubits able to simulate the effect of arbitrary target unitary operations $\hat{U}_t$
and an accompanying efficient compiling method.
For reasons related with the  compiling method,  we only consider circuits  consisting of $N=2^n$ circuit units
with identical two-qubit gate  architecture and placement of adjustable single qubit gates.
In \cite{Bu} has been shown that a $n$-qubit circuit able to simulate arbitrary $n$-qubit operations should consist by a minimum  number of
CNOTs equal to $\left(4^n-3n-1\right)/4$.
We use this limit to set a lower bound on   the number $N_{2q:uc}$ of two-qubit gates in the \textit{circuit unit} as  
\begin{equation}N_{2q:uc}\ge N^{min}_{2q:uc}=\left\lceil \left(4^n-3n-1\right)/ 2^{n+2}\right\rceil. \label{N2uc}\end{equation}
Concerning the number of single-qubit gates.
A generic $n$-qubit unitary matrix  $\hat{U}_t$ acting on $n$ qubits requires  $4^n-1$ real parameters for its description --in the case where the global phase is ignored as in this work. 
A  single-qubit operation  can be  parametrized as
\begin{equation}
 R\left(\phi_x,\phi_y,\phi_z\right)=\mathrm{exp}\left(i \phi_x \hat{\sigma}_x +i \phi_y \hat{\sigma}_y+i \phi_z \hat{\sigma}_z \right) \label{lo}
\end{equation}
where $\left\{\hat{\sigma}_j\right\}$ the Pauli matrices. Therefore  the minimum  number of 
  adjustable in all three angles, single-qubit  gates in the  \textit{circuit unit} is set to 
	\begin{equation}N_{1q:uc}\ge	N^{min}_{1q:uc}=\left\lceil 2^n/3\right\rceil.\label{N1uc}\end{equation} 
We denote   the totality of adjustable `local' parameters in the circuit as $\vec{\phi}$
and the subset of local parameters corresponding just to circuit unit as $\left\{\vec{\phi}\right\}$.
In addition, the total number of two-qubit gates and single-qubit gates, $N_{2q}$ and $N_{1q}$ respectively, can be calculated as $N_{2q}=2^n N_{2q:uc}$ and  
$N_{1q}=2^n N_{2q:uc}$. The structure of  circuits under study is presented schematically in Fig.~\ref{fig1}.

\section{The compiling method \label{CM}}
\begin{figure}[h] \includegraphics
[width=0.45\textwidth]
{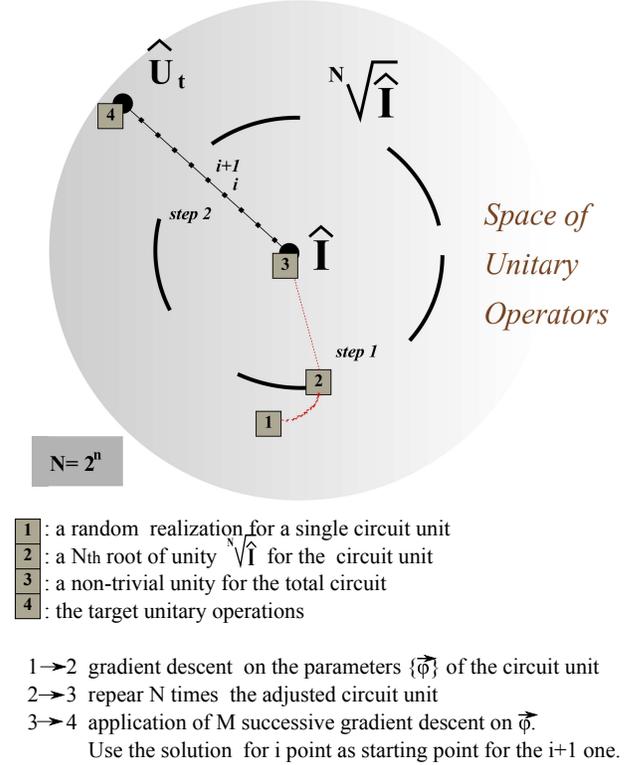}
\vspace{1cm}
\caption{ The steps of the compiling method. } \label{fig2}\end{figure}

In the following  we present the method for compiling a circuit taken from the class  defined  in Section~\ref{c}. Given a target unitary $\hat{U}_t$, the compiling consists of  tuning circuit's local parameters $\vec{\phi}$ such that this matches to  $\hat{U}_t$ at arbitrary high precision. 
The method consists   of two steps. The first step is performed just once for a given circuit and this only involves  the parameters $\left\{\vec{\phi}\right\}$ of a single circuit unit. At the end of first step one can also decide whether the circuit is compiling universal or not. The second step involves
all the parametric space $\vec{\phi}$ and should be performed every time a new target unitary $\hat{U}_t$ is given.
   The method is  schematically presented in Fig.~\ref{fig2}.

\subsection{First step: Engineering a non-trivial unity}

One works  at this step  with the circuit unit  and  aims to identify values for the  local
parameters $\left\{\vec{\phi}\right\}$  such as the spectrum of the circuit unit's unitary matrix   consists by all  $N$ roots of unity. Unitary operators with such spectrum are known  
as $N$th root of $\widehat{I}$ and we abbreviate these as  $\sqrt[N]{\widehat{I}}$.
This matching is computationally tractable  since there is a continuous manifold of $\sqrt[N]{\widehat{I}}$ operators and the matching  can be quantified   with the help of the  characteristic polynomial of the circuit unit's unitary matrix.

Let us  give more specific information on the aforementioned points. At this step one only  aims to control, via the local parameters $\left\{\vec{\phi}\right\}$,  the eigenvalues of the circuit's unitary matrix and not its eigenvectors. Therefore the solutions form a manifold  and starting from a random point a solution can be identified via a simple method such as gradient descent. Concerning the cost function to be used for the gradient descent method. In has been shown \cite{FE} that achieving the  $N$th root of unity for the circuit is (up to a global phase) equivalent
to vanishing the $N-1$ complex coefficients in the characteristic polynomial of the corresponding unitary matrix. To prove this, let us write the characteristic polynomial
for a unitary operation acting on $n$-qubits:
\begin{equation} x^N+\lambda_{N-1}\left(\left\{\vec{\phi}\right\}\right) x^{N-1}+\ldots +\lambda_{1}\left(\left\{\vec{\phi}\right\}\right) x+ e^{i \chi}~. \label{char} \end{equation}
It is easy now to see that the condition
$\sum_{j=1}^{N-1}\left| \lambda_j \left(\left\{\vec{\phi}\right\}\right)\right|=0 $  imposes  the eigenvalues of the unitary to be the $N$th roots of $-e^{i \chi}$. As a consequence the functional \begin{equation}\sum_{j=1}^{N-1}\left| \lambda_j\left(\left\{\vec{\phi}\right\}\right)\right|\label{obje} \end{equation}  is a convenient cost function for the  gradient method.

\vspace{1cm}

With the solution to this step being reached, we proceed by repeating the locally adjusted   circuit unit $N$ times, with $\vec{\phi}=\cup\left(\left\{\vec{\phi}\right\}\right)$. This series of locally adjusted  circuit units constitutes a non-trivial unity for the totality of the circuit (up to a phase $-e^{i \chi}$) and it is the starting point for the next step where the symmetry gets broken and
all the parameters  $\vec{\phi}$ are to be adjusted independently. There one needs to be able to `guide' the circuit  towards an arbitrary   direction in the space of unitary operators (see Fig.~\ref{fig2}) and the achievement of a non-trivial unity is a necessary condition for doing so. In addition  the architecture of the circuit should be efficient so that arbitrary targets in the space of unitary operators can be reached. The efficiency can be checked by setting few random target unitary operators in the neighborhood of unity and over-viewing
the averaging scaling of  gradient descent with the number of  steps, $K$. For compiling universality  the distance to a random target  should be exponentially
decreasing with the number of steps, as $\exp^{ -\gamma K}$ with $\gamma$ not negligible.

\subsection{Second Step: Reaching the target unitary via small steps }

Here, we present the algorithmic procedure for reaching a target unitary $\widehat{U}_t$ starting from the circuit's non-trivial unity.
For the purpose, it is useful to refer to a specific measure of distance between the target unitary $\widehat{U}_t$ and the unitary
corresponding to the adjusted circuit $\widehat{\widetilde{U}}_t$ and we choose to employ the simplest suggestion   \cite{Fowler, Kharti}:
\begin{equation} \mathcal{D}(\widehat{U}_t,\widehat{\widetilde{U}}_t)=1-\frac{1}{4^n}\left|\mathrm{tr}\left[\widehat{U}_t\widehat{\widetilde{U}}_t^{\dagger}\right]\right|^2 ~.\label{co2}\end{equation}

For every unitary matrix $\hat{U}$ one can identify a generator, a Hermitian matrix $\hat{H}$, such  as
\begin{equation}
\widehat{U}=\mathrm{e}^{\mathrm{i} \widehat{H}}.
\end{equation}
Given the target unitary, $\widehat{U}_t$,  we use its generator $\widehat{H}_t$ in order to  build a series of unitary operators 
\begin{equation}
\widehat{U}_t^{\left\{j,M\right\}}=\mathrm{e}^{\mathrm{i} \sqrt{\frac{j}{M}}\widehat{H}_t}, ~~ j=1,\ldots,M,
\label{M}\end{equation}
 where $\widehat{U}_t^{\left\{M,M\right\}}=\widehat{U}_t$.
A randomly picked unitary $\widehat{U}_t$ usually is at large distance  from $\widehat{I}$, i.e., $\mathcal{D}\left(\widehat{U}_t,\widehat{I}\right)\approx  1 $, while  $\mathcal{D}\left(\widehat{U}_t^{\left\{j,M\right\}},\widehat{I}\right)$
 is increasing approximately linearly with $j$ for $j<M/2$. 

The technique to reach $\hat{U}_t$ starting from the non-trivial unity consists of $M$ identical steps. One applies gradient descent
starting from $\widehat{I}$ to $\widehat{U}_t^{\left\{1,M\right\}}$ using as input the solution identified in step $1$ of the method. The output
 of the gradient descent is then used as input to the second step $\widehat{\widetilde{U}}_t^{\left\{1,M\right\}}\rightarrow\widehat{\widetilde{U}}_t^{\left\{2,M\right\}}$,
and so on. At each step $j<M$ the targeting distance $\mathcal{D}\left(\widehat{\widetilde{U}}_t^{\left\{j,M\right\}},\widehat{U}_t^{\left\{j,M\right\}}\right)$ can be fixed to a moderately high value as $0.01$. For the last step, $j=M$, one should put as objective the desired final accuracy. The method provides exact compiling since with an exponential decreasing gradient descent, one can approach as close as desired to the target unitary operation.

Let us note here that a quasi-Newton optimization method \cite{Vla} can be used as an alternative to the gradient descent method proposed here.
Finally, we mention  that the solution of the  intermediate  steps  can be used for tuning the circuit to simulate the time-evolution of a given Hamiltonian for a moderate number of qubits.

\section{Compiling universal CNOTs  circuits    \label{D}}
We numerically apply the methodology exhibited in Section~\ref{CM} in order to identify compiling
universal circuits constructed solely on the basis of CNOTs and single-qubit operations.
For $n=3$, $4$ and $5$ qubits we  identified  compiling universal circuits with
 $N_{2q:uc}= N^{min}_{2q:uc}$ and $N_{1q:uc}=\frac{3}{2}	N^{min}_{1q:uc}$, Eqs.(\ref{N2uc}),(\ref{N1uc}).
These results provide total counts on CNOTs gates, as $N_{2q}=16$, $64$ and $256$ respectively, which are very close to the minimum \cite{Bu} theoretical predictions: $14$, $61$, $252$. 

In the examples under study  we  also set  connectivity constrains between qubits. Like this we have been able
to see that the objective $N_{2q:uc}= N^{min}_{2q:uc}$  cannot be reached for all circuit architectural settings. We note thought that  the
 connectivity settings which are proven insufficient here  may result to compiling universal circuits when  $N_{2q:uc}> N^{min}_{2q:uc}$ or with the use of another compiling method.

The compiling universal circuits have in general different factors $\gamma$ describing the negative exponential scaling of the gradient descent with the number of steps
and therefore different \textit{compiling-time efficiency}. We have performed some preliminary comparison between  the compiling-time efficiency of different compiling universal circuits by setting  random targets,  fixing the final desired accuracy and counting the steps for reaching the later for each of circuits under study.

\subsubsection{ Three qubit circuits}
\begin{figure}[h] \includegraphics
[width=0.45\textwidth]
{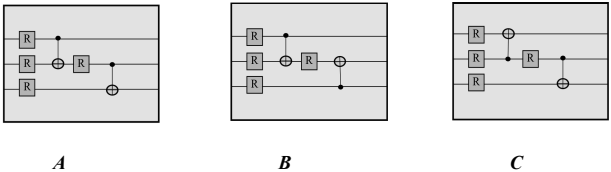}
\vspace{1cm}
\caption{Connectivity constrains between three qubits and the corresponding circuit units with $N_{2q:uc}= N^{min}_{2q:uc}=2$ for three qubits. All three circuits units  result to compiling universal circuits with total number of CNOTs $N_{2q}= 16$.} \label{fig3}\end{figure}

We have tested the three possible connectivity settings among the three qubits, see Fig.~\ref{fig3},
and we have concluded that all three   provide compiling universal circuits with $N_{2q:uc}= N^{min}_{2q:uc}=2$. 
 On the other hand the compiling-time efficiency of $A$ circuit is slightly higher than the one of circuit $B$, while for circuit $C$ the compiling-time efficiency 
is approximately $5$ times less than for $A$.
If  one more CNOT gate is added to circuit $A$, connecting qubit 3 to 1 and resulting to $N_{2q:uc}=3$,  the compiling-time efficiency  is increased approximately by a factor of  $3$ as compared to the one of  $A$. This effect has been observed for higher $n$ as well.

\subsubsection{ Four qubit circuits}
\begin{figure}[h] \includegraphics
[width=0.45\textwidth]
{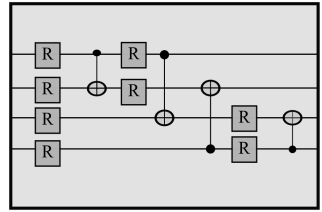}
\vspace{1cm}
\caption{We have tested  three different connectivity settings between four qubits and their corresponding unit circuits such that  $N_{2q:uc}= N^{min}_{2q:uc}$.
The unit circuit for $A$ connectivity is presented and the other two can be inferred. 
 $A$ and $C$ unit circuits generate compiling universal circuits
with $N_{2q}=64$ while $B$ does not.   } \label{fig4}\end{figure}

For four qubits we have not performed an exhaustive search but we have seen that
not every circuit unit with $N_{2q:uc}= N^{min}_{2q:uc}=4$ results to a compiling universal circuit, see for instance setting $B$ in Fig.~\ref{fig3}.  
The unit circuits  $A$ and $C$ in Fig.~\ref{fig4} result to compiling universal circuits that have approximately the same compiling-time efficiency.
We have observed that the order of placement of gates does not have an effect on compiling universality property or on the compiling-time efficiency.
 
\subsubsection{ Five qubit circuits}
For five qubits we have identified a few connectivity setting which can result in compiling universal circuits with
$N_{2q:uc}= N^{min}_{2q:uc}=8$.
Among the examples  we have studied are the IBM QX2 and IBM QX4 architectures. We found out that both architectures can provide compiling universal circuits with
$N_{2q}=256$, with
the IBM QX4 circuits having approximately double compiling-time efficiency than IBM QX2 circuits.

\section{Discussion \label{Di}}

We  propose  a method of  compiling   that permits to construct circuits of $2^n$-folded repetitive architecture
able to compile arbitrary unitary operations. This method works for a moderate number of qubits and
can be of potential use in the  field of quantum technologies. The compiling result can be considered as exact since
the accuracy is  increasing exponentially fast with the steps of the algorithm.

In this work we apply the method for identifying circuits composed exclusively by CNOT gates but the methodology is applicable
to circuits composed by any other type of two-qubit gates. As a future direction,  it would be interesting to apply the method in order to classify the 
efficiency of different types of two-qubit gates in the task  of building compiling universal circuits.

\section*{Acknowledgment}
AM is grateful to V. M. Akulin for providing  clarifications on technical points of his works \cite{FE, Vla}. This work has been financially supported by the Nazarbayev University ORAU grant ``Dissecting the collective dynamics of arrays of superconducting circuits and quantum metamaterials'' (no. SST2017031) and the MES RK state-targeted program BR$05236454$.



\begin{thebibliography}{99}
\bibitem {Kitaev} A. Y. Kitaev, \textsl{Quantum computations: algorithms and
error correction}, Russ. Math. Surv. \textbf{52}, 1191 (1997). 
\bibitem{QEP} M. A. Nielsen and I. L. Chuang, \textit{Quantum Computation and Quantum Information}, 
(Cambridge University Press, 2000).
\bibitem {Shen}A. Yu. Kitaev, A. Shen, and M. N. Vyalyi. \textit{Classical and quantum computation}, 1st edition. (American Mathematical Society, 2002). 

\bibitem {Deutsch2} D. Deutsch, A. Barenco, and A. Ekert, \textit{Universality in Quantum Computation}, Proc. R. Soc. London A \textbf{449}, 669 (1995). 
\bibitem{Harrow} A. W. Harrow, B. Recht, and I. L. Chuang, \textit{Efficient Discrete Approximations of Quantum Gates}, J. Math. Phys.  \textbf{43}, 4445 (2002). 


\bibitem {Nielsen}C. M. Dawson and M. A. Nielsen, \textit{ The Solovay-Kitaev algorithm}, Quant. Inf. Comp. \textbf{6}, 81 (2006). 
\bibitem {Fowler} A. G. Fowler, \textit{Constructing arbitrary Steane code single logical qubit fault-tolerant gates}, Quant. Inf. Comp. \textbf{11}, 867 (2011). 
\bibitem {Svore} A. Bocharov and K. M. Svore, \textit{Resource-optimal single-qubit quantum circuits}, Phys. Rev. Lett. \textbf{109}, 190501 (2012). 
 \bibitem{Booth} J. Booth Jr, \textit{Quantum compiler optimizations}, arXiv:1206.3348 (2012).
\bibitem{SvoreII} A. Bocharov, Y. Gurevich, K. M. Svore, \textit{Efficient Decomposition of Single-Qubit Gates into V Basis Circuits}, Phys. Rev. A \textbf{88}, 012313 (2013).
\bibitem {Horsman}T. T. Pham, R. Van Meter, and C. Horsman, \textit{Optimization of the Solovay-Kitaev algorithm}, Phys. Rev. A \textbf{87}, 052332 (2013). 
\bibitem {Mosca} V. Kliuchnikov, D. Maslov, and M. Mosca, \textit{Asymptotically optimal approximation of single qubit unitaries by Cliffordand T circuits using a constant number of ancillary qubits}, Phys. Rev. Lett. \textbf{110}, 190502 (2013). 
 \bibitem{mand} Y. Zhiyenbayev, V. M. Akulin, and A. Mandilara, \textit{Quantum compiling with diffusive sets of gates}, Phys. Rev. A \textbf{98}, 012325 (2018).



\bibitem {Deutsch} A. Barenco, C. H. Bennett, R. Cleve, D. P. DiVincenzo,
N. Margolus, P. Shor, T. Sleator, J. A. Smolin, and H. Weinfurter, \textit{Elementary gates for quantum computation}, Phys.Rev. A \textbf{52}, 3457 (1995).
\bibitem{Salomaa1} J. J. Vartiainen, M. M\"{o}tt\"{o}nen, M. M. Salomaa, \textit{Efficient decomposition of quantum gates}, Phys. Rev. Lett. \textbf{92}, 177902 (2004).
\bibitem{Salomaa3} M. M\"{o}tt\"{o}nen, J. J. Vartiainen, V. Bergholm, and M. M. Salomaa, \textit{Quantum Circuits for General Multiqubit Gates}, 
Phys. Rev. Lett. \textbf{93}, 130502 (2004).
\bibitem{Vidal} G. Vidal and C. M. Dawson, \textsl{Universal quantum circuit for two-qubit transformations with three controlled-NOT gates}, Phys. Rev. A 69, 010301(R) (2004).
\bibitem{Markov} V. V. Shende, S. S. Bullock, I. L. Markov, \textsl{Synthesis of Quantum Logic Circuits}, IEEE Trans. on Computer-Aided Design  25, 1000 (2006).


\bibitem{Sanders} S. Raeisi, N. Wiebe, and B. C. Sanders, \textit{Quantum-circuit design for efficient simulations of many-body quantum dynamics}, New J. Phys. \textbf{14}, 103017 (2012).
\bibitem{Reck} M. Reck, A. Zeilinger, H. J. Bernstein, and P. Bertani, \textit{Experimental realization of any discrete unitary operator}, Phys. Rev. Lett. \textbf{73}, 58 (1994).

\bibitem{Kharti} S. Khatri, R. La Rose, A. Poremba, L. Cincio, A. T. Sornborger, and P. J. Coles, \textit{Quantum-assisted quantum compiling}, Quantum \textbf{3}, 140 (2019).

\bibitem{Ar} T. Peng, A. Harrow, M. Ozols, X. Wu, \textit{Simulating Large Quantum Circuits on a Small Quantum Computer}, Phys. Rev. Lett. \textbf{125}, 150504 (2020).

\bibitem{San} E. Zahedinejad, J. Ghosh and B. C. Sanders, \textit{High-Fidelity Single-Shot Toffoli Gate via Quantum Control}, Phys. Rev. Lett. \textbf{114}, 200502 (2015).


\bibitem{Bu} V. V. Shende, I. L.Markov, and S. S. Bullock, \textit{Minimal universal two-qubit controlled-NOT-based circuits},  Phys. Rev. A \textbf{69},062321, (2004).





 






\bibitem{FE}  G. Harel and V. M. Akulin,  \textit{Complete Control of Hamiltonian Quantum Systems: Engineering of Floquet Evolution},   Phys. Rev. Lett.   \textbf{82}, 1    (1999).





\bibitem{Vla} V. M. Akulin, V. Gershkovich, and G. Harel, \textsl{Nonholonomic quantum devices}, Phys. Rev. A \textbf{64}, 012308 (2001).

























\end{thebibliography}
\end{document}